\documentstyle[12pt,amstex]{article}

\newfont{\bit}{cmbxti10 scaled 1728}

\def\RR{ I \hspace*{-0.8ex} R }

\def\oh{\hat{\omega}}
\def\ot{\tilde{\omega}}
\def\at{\tilde{a}}

\begin{document}
\renewcommand{\thefootnote}{\fnsymbol{footnote}}
\newpage
\pagestyle{empty}
\begin{center}
{\LARGE {A Note on the  Symmetries \\
 of the\\
Gravitational Field of a Massless Particle\\
}}

\vspace{2cm}
{\large
 Peter C. AICHELBURG
 \footnote[1]{ e-mail: pcaich @@ pap.univie.ac.at}
\footnote[4]{This work was also supported in part by the FUNDACION FEDERICO}
}\\
{\em
 Institut f\"ur Theoretische Physik, Universit\"at Wien\\
 Boltzmanngasse 5, A - 1090 Wien, AUSTRIA
 }\\[.5cm]
{\em and}\\[.5cm]
{\large Herbert BALASIN
\footnote[2]{e-mail: hbalasin @@ ecxph.tuwien.ac.at}
\footnote[3]{supported by the APART-program of the Austrian Academy
 of Sciences}
}\\
{\em
 Institut f\"ur Theoretische Physik, Technische Universit\"at Wien\\
 Wiedner Hauptstra{\ss}e 8--10, A - 1040 Wien, AUSTRIA
}\\[.5cm]
\end{center}
\vspace{2cm}

\begin{abstract}
It is shown that the metric of a massless particle obtained from
boosting the Schwarzschild metric to the velocity of light, has four
Killing vectors corresponding to an $E(2)\times \RR$ symmetry-group.
This is in agreement with the expectations based on flat-space
kinematics but is in contrast to previous statements in the
literature \cite{Schueck}. Moreover, it also goes beyond the general
Jordan-Ehlers-Kundt-(JEK)-classification of gravitational pp-waves as given
in \cite{JEK}.

\end{abstract}
\vfill

\rightline{UWThPh -- 1994 -- 27}
\rightline{TUW 94 -- 12}
\rightline{July 1994}

\renewcommand{\thefootnote}{\arabic{footnote}}
\setcounter{footnote}{0}
\newpage
\pagebreak
\pagenumbering{arabic}
\pagestyle{plain}

Several years ago R. Sexl and one of the authors (P.C.A.) obtained
a gravitational impulsive plane-fronted wave with parallel rays (pp-wave)
by boosting the Schwarzschild metric to the singular limit \cite{AS}.
Since then this metric\footnote{denoted AS-metric by some authors}
 has been widely used to discuss ultra-relativistic
scattering processes in both the classical and semiclassical context
\cite{tHooft,Verl}.
Moreover, the method of boosting was applied  also to other configurations
like the Kerr or the Reissner-Nordstr{\o}m metric \cite{LoSa,Ferr}.
In flat spacetime the trajectory of a free particle of zero rest mass
is described by its initial position and 4-momentum which is null.
The subgroup of the Poincare group that leaves the trajectory
invariant is isomorphic
to  $E(2)\times \RR$. Its generators act in the null-hyperplane
orthogonal to the given vector. On physical grounds one would therefore
expect that the gravitational  field generated by a massless particle
shares the same symmetries. Already in 1962, in a now classical paper
\cite{JEK}, Ehlers et al gave a complete classification of symmetries
for pp-waves together  with their  standard form  for each symmetry. A
comparison of this classification with the metric of a massless particle
tells one that the metric should have only two instead of the expected
four Killing vectors, namely the translation along the given lightlike
propagation direction and the corresponding space rotation.
This contradiction was emphasized in a paper by Sch\"ucker \cite{Schueck}
who argued that the full symmetry implies the presence of infinite mass
density which general relativity does not allow for.
In this note we show that the symmetry of the field in question indeed
admits the expected four Killing vectors. The above mentioned
contradiction is resolved by noting that the metric, which is an impulsive
gravitational wave, has in its standard form a distributional metric
coefficient. Thus, allowing for distributions a careful analysis of
the Killing equations shows the presence of the full symmetry. It is
also straightforward to see that the JEK-classification is based on
regular functions only.
\par

Our approach to derive the Killing vectors of the AS-geometry will
start from its unboosted counterpart, the Schwarzschild geometry.
Since both geometries, Schwarzschild and AS, belong to the Kerr-Schild
class they admit a decomposition
$$
g_{ab} = \eta_{ab} + f\; k_a k_b,
$$
where $\eta$ denotes the flat part of the decomposition, $k$ a
geodetic null vector-field and $f$ a scalar function. The existence of
the flat background-part in the decomposition gives a natural meaning to
the concept of boosts as its isometries. Moreover this geometrical aspect
will turn out to be of great calculational advantage. Choosing Kerr-Schild
coordinates $x^a$ for the Schwarzschild metric brings
$\eta$ into its standard form and $f$ and $k$ become respectively
\begin{equation}\label{SS}
f=\frac{2m}{r}\qquad k^a = {1 \choose {\textstyle\frac{\; x^i}{r}}}
\end{equation}
where $x^i$ denotes spatial cartesian coordinates according to the
(flat) Min\-kow\-ski metric and $r$ its corresponding radial distance.
With respect to these coordinates the Killing vectors of (\ref{SS})
corresponding to staticity and spherical symmetry can be written as
\begin{align}\label{SSKill}
&\xi_t^a = (\partial_t)^a \\
&\xi_{\oh}^a = \oh^a{}_b x^b\qquad \oh^a{}_b \xi^b_t=0\quad
\oh^{ab} = -\oh^{ba},\nonumber
\end{align}
where $\oh$ denotes an infinitesimal rotation matrix.
The index-raising in the last line of (\ref{SSKill}) was performed with
the help of $\eta$, which will be implicitly assumed in the following
unless stated otherwise.
Interpreting the Kerr-Schild coordinates as referring to a Lorentz-frame
associated with
an observer that is at rest asymptotically, allows us to find the
form of the Killing vectors with respect to a boosted asymptotic
observer
\begin{align}
&\xi^a_t = \frac{1}{m} P^a,  \\
&\xi^a_{\ot} = \ot^a{}_b x^b, \nonumber \\
&\xi^a_{\at} = \frac{1}{m} ((\at x)Q^a - (Qx)\at^a ), \nonumber\\
&\oh^a{}_b = \ot^a{}_b + \frac{1}{m}(Q^a \at_b - \at^a Q_b ),\nonumber
\end{align}
where $P$ and $Q$ denote the timelike and spacelike vectors
spanning the boost 2-plane and vectors in the orthogonal 2-plane
are tagged by a tilde (i.e. $P^a\at_a =Q^a\at_a=0$ ). $P$ and $Q$
are conveniently normalized to $\pm m^2$ and tend to a null vector $p$
in the limit $m\to 0$.
%
%
To take $m \to 0$, we multiply $\xi_t$ and $\xi_{\at}$ with
$m$ thereby obtaining the limit
\begin{align}\label{ASKill}
&(m\xi_t^a) \to p^a =: \zeta_t^a,\\
&\xi^a_{\ot} \to \ot^a{}_b x^b =: \zeta^a_{\ot},\nonumber\\
&(m\xi^a_{\at}) \to (\at x)p^a - (px)\at^a  =: \zeta_{\at}^a.\nonumber
\end{align}
{}From their mutual commutation relations
\begin{alignat*}4
&\left[ \zeta_t, \zeta_{\ot} \right ] = 0 &\hspace{2cm} &\left[ \zeta_t,
\zeta_{\at} \right ] =0&\hspace{4cm} & \\
&\left[ \zeta_{\at},\zeta_{\tilde{a}'} \right ] = 0&\hspace{2cm} &
\left[ \zeta_{\ot}, \zeta_{\at} \right ] = \zeta_{\ot\at} &  \hspace{4cm} &
\end{alignat*}
one sees that the $\zeta$ form a representation of the Lie-algebra
of $E(2)\times\RR$.

\noindent
The corresponding limit of the Schwarzschild geometry
\begin{equation}\label{ASmet}
g_{ab} = \eta_{ab} - 8\delta (px)\log\rho\;p_a p_b,
\end{equation}
may either be obtained by employing a ''singular'' boost \cite{AS}
or by solving the Einstein equations with a boosted source \cite{BaNa}.
The result is a pp-wave with profile function $f=-8\delta(px)\log\rho$,
where $\rho$ denotes the radial distance in the orthogonal 2-plane.
It is now easy to verify that all the $\zeta$ of (\ref{ASKill}) have the
Killing-property
\begin{align}\label{Killeq}
L_\zeta g_{ab} &= (\zeta\partial)f\; p_ap_b + \partial_a\zeta_b +
\partial_b\zeta_a + f (p_a\partial_b(p\zeta) + p_b\partial_a(p\zeta)) = \\
&= (\zeta\partial)f\; p_ap_b + \partial_a\zeta_b +
\partial_b\zeta_a =0\nonumber,
\end{align}
where $\partial$ is the covariant derivative with respect to $\eta$.
Let us however, for the sake of explicitness, exhibit the only non-trivial
term of (\ref{Killeq}) relative to $\zeta_{\at}$
$$
(\zeta_{\at}\partial)f = -8(\at x)p^2\delta '(px) +
8(px)\,\delta(px)\frac{1}{\rho}e_\rho^a \at_a,
$$
which vanishes due to the product $px\,\delta (px)$ and the fact
that $p$ is null.

This shows that the Killing vectors of the Schwarzschild metric
survive the ultrarelativistic boost in the sense that the resulting
vector fields are Killing with respect to the boosted metric.
Comparing (\ref{ASmet}) with the general classification of Killing
vectors for pp-waves \cite{JEK} one would only expect the presence of
$\zeta_t$ and $\zeta_{\ot}$. The fact that the 2-parameter family of
vector fields $\zeta_{\at}$ is also Killing
depends crucially upon the presence of the $\delta(px)$-factor in $f$,
which also explains the discrepancy with \cite{JEK}, where no
distributional profile functions were considered. The implications of
this aspect for the general classification of pp-waves will be
investigated in a forthcoming publication.
\vfill
\noindent
{\em Acknowledgement: }The authors wish to thank H.~Urbantke for a helpful
comment. One of the authors (P.C.A.) thanks A.~Ashtekar for
a discussion on this problem during a meeting in Cocoyoc, Mexico back in 1990.

\vfill
\end{document}